\documentclass[%
aps,%
prb,%
preprint,
showpacs,%
preprintnumbers,%
byrevtex,%
floatfix%
]{revtex4}

\usepackage{epsf}
\usepackage{ifthen}
\usepackage[usenames]{color}
\usepackage{amssymb}
\usepackage{amsfonts}
\usepackage{amsmath}
\usepackage[dvips]{graphicx}

\usepackage[dvips, colorlinks=true, bookmarks, pdftitle={An example PDF file from LaTeX},
 pdfauthor={Gianaurelio Cuniberti}, pdfsubject={From LaTeX to PDF}, pdfkeywords={PDF, LaTeX, hyperlinks, hyperref}]{hyperref}

\def\ii{\textrm{i}\,}

\def\ie{\textit{i.e.}}

\def\eg{\textit{e.g.}}

\begin{document}
\date{October, 01 2009}
\title{Structural fluctuations and quantum transport through DNA molecular wires: a combined molecular dynamics and model Hamiltonian approach}

\author{R. Guti{\'e}rrez}
\affiliation{Institute for Materials Science and Max Bergmann Center of Biomaterials, Dresden University of Technology, 01062, Dresden, Germany}
\author{R. Caetano}
\affiliation{Instituto de Fisica, Universidade Federal de Alagoas, Maceio, AL 57072-970, Brazil}
\author{P. B. Woiczikowski}
\affiliation{Institute for Physical Chemistry, University Karlsruhe, 76131 Karlsruhe, Germany}  
\author{T. Kubar}
\affiliation{Institute for Physical and Theoretical Chemistry, Technical University Braunschweig,38106, Braunschweig, Germany}
\author{M. Elstner}
\affiliation{Institute for Physical Chemistry, University Karlsruhe, 76131 Karlsruhe, Germany}  
\author{G. Cuniberti}
\affiliation{Institute for Materials Science and Max Bergmann Center of Biomaterials, Dresden University of Technology, 01062, Dresden, Germany}

\begin{abstract}
Charge transport through a short DNA oligomer (Dickerson dodecamer) in presence of structural fluctuations is investigated using a  hybrid computational methodology based on a combination of quantum mechanical electronic structure calculations and classical molecular dynamics simulations with a model Hamiltonian approach. 
Based on a fragment orbital description, the DNA electronic structure can be coarse-grained in a very efficient way. 
The influence of dynamical fluctuations arising either from the
solvent fluctuations or from  base-pair vibrational modes can be taken into account in a straightforward way
 through  time series of the effective DNA electronic parameters, evaluated  at snapshots along the MD trajectory. We show that charge transport can be promoted through the coupling to solvent  fluctuations, which gate the onsite energies along the DNA wire. 
\end{abstract}


\maketitle

\section{Introduction}

The electrical response of DNA oligomers to applied voltages is a highly topical issue which has attracted the attention of scientists belongig to different research communities. The variability of experimental results is reflected in DNA being predicted to be an insulator,~\cite{braun98} a semiconductor,~\cite{porath00,cohen05} or a metallic-like system.~\cite{tao04,yoo01} This fact hints not only at the difficulties encountered in carrying out well-controlled single-molecule experiments, but also at the dramatic sensitivity of charge migration through DNA molecules to
intrinsic, system-related or extrinsic, set-up mediated factors: the specific base-pair sequence, internal vibrational excitations, solvent fluctuations, and the electrode-molecule interface topology, among others.  As a result, the theoretical modelling of DNA quantum transport remains a very challenging issue that has been approached from many different sides, see \eg, Refs.~\onlinecite{porath04,starikow05,levy05,tapash07} for recent reviews. While most of the models originally used started from a static picture of the DNA structure,~\cite{roche03,gio02,shih:018105,guo:035115} it has become meanwhile clearer that charge migration through DNA oligomers attached to electrodes may only be understood in the context of a {\textit {dynamical approach}}.~\cite{hennig04,starikov07,CramerT._jp071618z,GrozemaFerdinandC._ja078162j,GrozemaF_ja001497f,troisi02,gutierrez2009,gutierrez2009a} Hole transfer experiments~\cite{meggers98,meggers98a,treadway02,tb98,wan99} had already  hinted at the strong influence of DNA structural fluctuations in supporting or hindering charge propagation. Hence, it seems natural to expect that dynamical effects would also play a determining role in charge transport processes for molecules contacted by electrodes. A realistic inclusion of the influence of dynamical effects onto the transport properties can however only be achieved   via hybrid methodologies  combining a reliable description of the biomolecular dynamics and electronic structure with  quantum transport calculations. {\textit{Ab initio}} calculations for static biomoecular structures can provide a very valuable starting point for the parametrization of model Hamiltonians;~\cite{bixon00,voityuk:115101,voityuk:034903,mehrez05,difelice02,felice02,felice04,star2003,ladik:105101} however, a full first-principle  treatment of both dynamics and electronic structure  lies outside the capabilities of state-of-the-art methodologies.

In this paper, we will ellaborate on a recent study on homogeneous DNA sequences~\cite{gutierrez2009} by addressing in detail some methodological issues. Our focus will be on the so called Dickerson dodecamer~\cite{dickerson} with the sequence $3^{'}-$GC{{GCTTAACG}}GC$-5^{'}$ and for which the effect of the dynamical fluctuations becomes very clear. Our approach combines classical molecular dynamics (MD) simulations with electronic structure calculations to provide a realistic starting point for the description of the influence of structural fluctuations onto the electronic structure of a biomolecule. Further, the information drawn from such calculations be used to formulate low-dimensional effective Hamiltonians to describe charge transport. A central point is the use of the concept of fragment orbitals~\cite{SGG2005,markus1,markus2,song} which allows for a very efficient and flexible mapping of the electronic structure of a complex system onto a much simpler effective model. Taking a single base pair as a fragment and considering only one fragment orbital per base pair,  we end up in a tight-binding Hamiltonian for a linear chain where both onsite energies $\epsilon_{j}(t)$ and electronic coupling terms $V_{j,j+1}(t)$ are time-dependent variables:
\begin{eqnarray}
\label{eq1}
H=\sum_{j} \epsilon_{j}(t) d^{\dagger}_{j}d_{j} + \sum_{j} V_{j,j+1}(t)\,(d^{\dagger}_{j}d_{j+1} + {\textrm h.c.}).
\end{eqnarray}
The dynamical information provided in this way builds the starting point of our treatment of quantum transport through biomolecular wires. 
In the next Section, we briefly describe the computational methodology used to obtain the effective electronic parameters of the model Hamiltonian, which will be then introduced in Sec.~{\textbf{III}}, where we also illustrate how to relate the Hamiltonian of Eq.~(\ref{eq1}) to a different model describing the coupling of a time-independent electronic system to a bosonic bath. Further, expressions for the electrical current as well as the relation between the auto-correlation function of the onsite energy fluctuations and the spectral density of the bosonic bath will be derived. Finally, we discuss in Sec.~{\textbf{IV}} the transport properties of the Dickerson dodecamer in vacuum and in presence of a solvent. We stress that in contrast to other models which explicitly contain the coupling to vibrational excitations or to an environment~\cite{gutierrez06,gmc05a,gmc05b,schmidt:115125,schmidt:165337} at the price of introducing several free parameters, our methodology  potentially contains the full dynamical complexity of the biomolecule as obtained from the MD simulations. One main advantage of our approach is the possibilty to progressively improve the degree of coarse-graining by an appropriate re-definition of the molecular fragments. 

\section{Molecular dynamics and model Hamiltonian formulation}
\subsection{Computational methodology}
We will first give an overview of the  fragment-orbital  approach used in our computations; further details can be found elsewhere.~\cite{markus1,markus2,VSR2004}

\begin{figure}[t]
\centerline{
\epsfclipon
\includegraphics[width=0.99\linewidth]{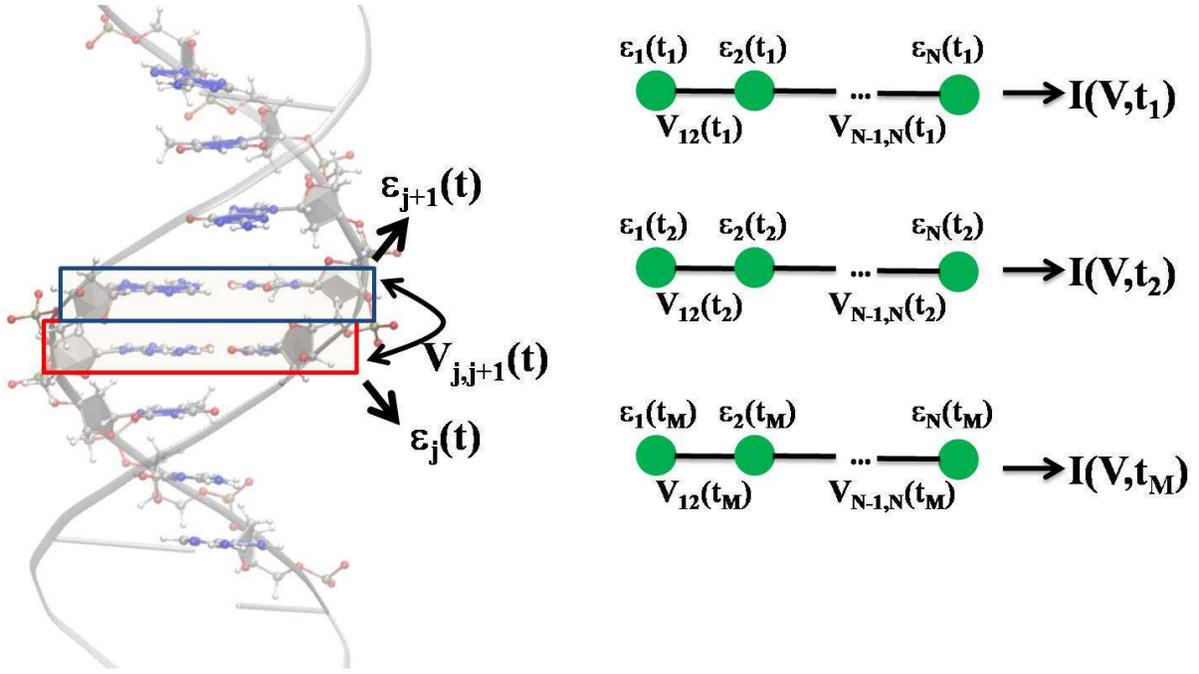}%
\epsfclipoff
}
\caption{\label{fig:FOM}%
Left panel: Schematic representation of the fragment orbital method used to perform a coarse-graining of the DNA electronic structure. A fragment consists of a single base pair (not including the sugar phosphate backbones). As explained in the text, the hopping matrix elements $V_{j,j+1}$ between nearest-neighbor fragments are compued using the molecular orbital basis of the isolated base pairs. These calculations are then carried out at snapshots along the molecular dynamics trajectory hence leading to time dependent electronic structure parameters. By keeping only one relevant orbital per fragment, the electronic structure can be mapped onto that of a linear chain (right panel). Transport observables can be computed at each simulation time step. Alternatively, the time dependence o the electronic structure (related to structural and solvent fluctuations) can be transferred to a bosonic bath as done in this paper. 
}
\end{figure}

The core of our method is  based on a combination of a charge self-consistent density-functional
parametrized tight-binding approach (SCC-DFTB) and the  fragment orbital concept;
\cite{VSR2004} both are used to  compute the electronic parameters $\epsilon_{j}(t)$ and  $V_{j,j+1}(t)$ for the effective tight
binding model introduced in  Eq.~(\ref{eq1}) in a very efficient way, see Fig.~\ref{fig:FOM} for a schematic representation. The  $V_{j,j+1}$ are calculated using the
 highest occupied molecular orbital $\Phi_i$ computed for isolated bases as $V_{j,j+1} = \left\langle \Phi_j|H|\Phi_{j+1}\right\rangle ,$  where
the $\Phi_j$'s can be expanded in a valence atomic orbital  basis $\eta_{\mu}$ on a given fragment:
$\Phi_j = \sum_{\mu} c_{\mu}^j \eta_{\mu}$.
The $ c_{\mu}^j$ are obtained from calculations on isolated bases and stored for subsequent use to
calculate $V_{j,j+1}$. Therefore, one can write:
\begin{eqnarray}
V_{j,j+1} =  \sum_{\mu}  \sum_{\nu} c_{\mu}^j c_{\nu}^{j+1}  \left\langle \eta_{\mu}|H|\eta_{\nu}\right\rangle .
\label{T}
\end{eqnarray}
The Hamilton matrix in the atomic orbital basis $H_{\mu \nu}=\left\langle \eta_{\mu}|{H}|\eta_{\nu}\right\rangle $   evaluated using the
SCC-DFTB Hamiltonian matrix is pre-calculated and stored, thus making this step
extremely efficient, {\textit{i.e.}},  it can be calculated for geometry snapshots generated by a classical
molecular dynamics  simulation even for several nanoseconds. Additionally, the minimal LCAO basis set used in the standard 
SCC-DFTB code has been optimized for the calculation of the hopping matrix elements,~\cite{markus1,markus2}
and the results are in very good agreement with other approaches.~\cite{VSR2004,voityuk:115101}
Concerning now the coupling to the solvent, a hybrid quantum mechanics/molecular mechanics (QM/MM) approach has been used, implemented in  the SCC-DFTB code,~\cite{CuiQ._jp0029109} leading to the following Hamiltonian matrix in the valence atomic orbital basis $\eta_{\nu}$:
\begin{eqnarray}
H_{\mu \nu}^{\textrm{QM/MM}}= H_{\mu \nu}^{\textrm{SCC-DFTB}} + \frac{1}{2}  S_{\mu  \nu} \left\lbrace \sum_{\delta} Q_{\delta} \left(
 \frac{1}{R_{\alpha \delta}} + \frac{1}{R_{\beta \delta}}  \right)+\sum\limits_A Q_A \left(\frac{1}{r_{A\alpha}}+ \frac{1}{r_{A\beta}}\right)\right\rbrace  .
 \label{QMMM}
%
 \end{eqnarray}

Here, $Q_{\delta}$ are the Mulliken charges of the quantum-mechanical region and the $Q_A$ are the charges in the MM region (backbones, counter-ions, and water), $S_{\mu  \nu}$ is the atomic orbital overlap
matrix, $H_{\mu \nu}^{\textrm{SCC-DFTB}}$ is the corresponding zero-order Hamiltonian matrix, and
$R_{\alpha \delta}$ is the distance between the DNA atom where the AO orbital $\eta_{\mu}$ is
located and  the MM atom in the solvent with charge $Q_{\delta}$. The last term explicitly takes into account the coupling to the environment (solvent). The $\epsilon_j=V_{jj}$ and $V_{j,j\pm 1}$
from Eq.~(\ref{T}) using Eq.~(\ref{QMMM}) can now be calculated along
the MD trajectories~\cite{markus2,gutierrez2009a}. The off diagonal matrix elements strongly depend
 on structural fluctuations of the DNA base pairs, but they are only weakly affected by the solvent dynamics, the
 opposite holding for the onsite energies which are considerably modified by solvent fluctuations.~\cite{markus2,gutierrez2009a} The
    Fourier transform of the onsite energies auto-correlation function provides
  information about the spectral ranges which are more strongly contributing to the fluctuations of the
  electronic parameters, see also the next sections.  Thus, we have found that the
apparently most important modes are located around 1600 cm$^1$  corresponding to  a base
skeleton mode, and at 800 cm$^{-1}$ related to the water modes.~\cite{markus2} Both contributions modulate the onsite
energies significantly on a short time scale and a long time scale of about 1 ps, respectively.

\subsection{Model Hamiltonian for electronic transport}

Using Eq.~(\ref{eq1}) directly for quantum transport calculations may mask to some degree  different
contributions (solvent, base dynamics) to charge propagation through the DNA $\pi$-stack. Moreover,
since Eq.~(\ref{eq1}) contains random variables through the time series, we are confronted with the problem of dealing with
 charge transport in an stochastic Hamiltonian. This is a more complex task which has been addressed \eg, in the context of exciton transport~\cite{PhysRevB.31.2479,PhysRevB.31.2430,haken72,reineker75} but also to some degree in electron transfer theories.~\cite{SpirosS.Skourtis03082005,GudowskaNowak1996115,goychuk:4937} Here, we adopt
 a different point of view and formulate a model Hamiltonian, where the relevant electronic system,
 in this case the fragment orbital-derived effective DNA electronic system, is coupled explicitly to a bosonic bath.
 The latter will encode through its spectral density the dynamical information drawn from the MD
 simulations on internal base dynamics as well as solvent fluctuations. The Hamiltonian can be written
 in the following way:
\begin{eqnarray}
\label{eqHam}
H&=&\sum_{j} \left\langle \epsilon_{j}\right\rangle_{t}  d^{\dagger}_{j} d_{j} -\sum_{j}
\left\langle V_{j,j+1}\right\rangle_{t}   ( d^{\dagger}_{j} d_{j+1} + {\rm h.c.}) \\
&+&  H_{\rm bath}+ H_{\rm el-bath}+H_{\rm tunnel} + H_{\rm leads} \nonumber \\
H_{\rm bath}&=& \sum_{\alpha} \Omega_{\alpha} B^{\dagger}_{\alpha}B_{\alpha} \nonumber \\
H_{\rm el-bath}&=& \sum_{\alpha,j} \lambda_{\alpha} d^{\dagger}_{j} d_{j} (B_{\alpha}+B^{\dagger}_{\alpha})
\nonumber \\
H_{\rm tunnel}&=& \sum_{{\bf k},s,j} \left( t_{{\bf k}s,j} c^{\dagger}_{{\bf k}s}d_{j} + {\rm h.c.} \right)
\nonumber \\
H_{\rm leads}&=&\sum_{{\bf k},s} \epsilon_{{\bf k}s} c^{\dagger}_{{\bf k}s} c_{{\bf k}s} \nonumber
\label{polaron}
\end{eqnarray}
The  time averages (over the corresponding time series) of the electronic parameters
$\left\langle \epsilon_{j}\right\rangle_{t} $ and $\left\langle V_{j,j+1}\right\rangle_{t} $
have been split off to provide a zero-order Hamiltonian which contains dynamical effects in a
mean-field-like level.  The effect of the fluctuations around these averages is hidden in the
vibrational bath, which is  assumed to be a collection of a large ($N\to\infty$) number of harmonic
oscillators in thermal equilibrium at temperature $k_{\rm B}T$. The bath will be characterized  by
a spectral density $J(\omega)$ which can also be extracted from the MD simulations, as shown below
and in Ref.~\cite{PhysRevE.65.031919}. Since we are interested in calculating the electrical
response of the system, the charge-bath model has to also include the coupling of the system to
 electronic reservoirs (electrodes). The coupling to the electrodes will be treated in a standard way,
 using a tunneling Hamiltonian $H_{\textrm {tunnel}}$ which describes the coupling to the $s$-lead with
 $s$= left (L) or right (R). Later on,  the so called wide-band limit will be introduced
 (the corresponding electrode self-energies are purely imaginary and energy-independent),
 thus reducing the electrode-DNA coupling to a single parameter.

\begin{figure}[t]
\centerline{
\epsfclipon
\includegraphics[width=0.99\linewidth]{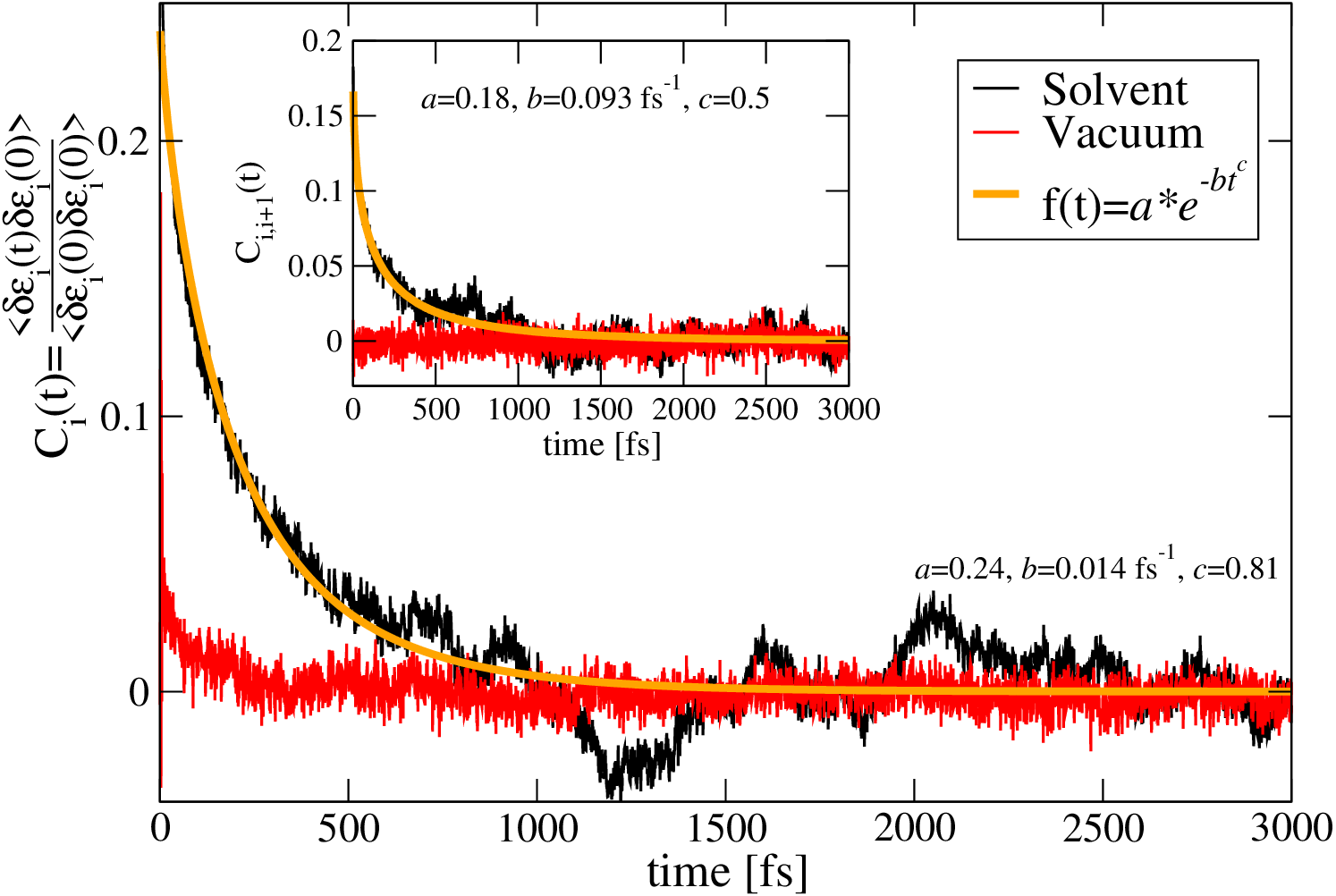}%
\epsfclipoff
}
\caption{\label{fig:FFT}%
Normalized auto-correlation function~\cite{PhysRevE.65.031919}
$C_{{ {i}}}(t)=\left\langle \delta\epsilon_{i}(t) \delta\epsilon_{i}(0)
\right\rangle/\left\langle \delta\epsilon^{2}_{i}\right\rangle$
 and averaged nearest-neighbor correlation function
 $C_{{\textrm {i,i+1}}}(t)=\left\langle \delta\epsilon_{i}(t) \delta\epsilon_{i+1}(0) \right\rangle /\left\langle \delta\epsilon_{i}\delta\epsilon_{i+1} \right\rangle$ (inset)
  of the onsite energy fluctuations.  The solid lines are fits
  to stretched exponentials which suggests the existence
  of different time scales, a typical situation  in the dynamics of bio-molecules.
  On average, the decay of $C_{{ {i,i+1}}}(t)$ occurs on a much shorter time scale than
  that of $C_{{\textrm {i}}}(t)$, so that our model will only include in a first approximation  local fluctuations, see Eq.~(\ref{eq1}).
}
\end{figure}

The previous  model relies on some basic assumptions that can be substantiated by the results of
the MD simulations:~\cite{markus2} (i) The complex DNA dynamics can be well mimic within the harmonic
 approximation by using a continuous vibrational spectrum; (ii) The  simulations show that the local
  onsite energy fluctuations are much stronger in presence of a solvent than those of the electronic
  hopping integrals (see also the end of the previous section), so that we assume that the bath is
  coupled only diagonally to the charge density fluctuations; (iii) Fluctuations on different sites
  display rather similar statistical properties, so that the charge-bath coupling $\lambda_{\alpha}$
  is taken to be independent of the site $j$. This latter approximation can be lifted by introducing additional site-nonlocal spectral densities $J_{j,j+1}(\omega)$; this however would make the theory more involved and less transparent. In Fig.~\ref{fig:FFT} we show typical normalized auto-correlation
  functions of the onsite energy  fluctuations for the Dickerson dodecamer in both  solvent and vacuum conditions as well as
   one case of off-diagonal correlations between  nearest-neighbor site energies (inset). Also shown are  fits to stretched exponentials, which are in general equivalent to a sum of simple exponential functions. This suggests the presence of different time scales and the non-trivial time-dependence of the fluctuation dynamics. The fits  become obviously 
    less accurate at long times due to the reduced number of sampled data points with increasing time
    (the oscillatory behavior becomes stronger). From the figure we first see that the off-diagonal correlations decay on a much shorter time scale as the local ones, so that on a first approximation their neglection can be justified; further the decay of the correlations for the vacuum simulations is considerably much faster than in a solvent indicating the strong influence of the latter in gating the electronic structure of the biomolecule.  
   The corresponding correlation functions for the hopping integrals $V_{j,j+1}(t)$ (not shown) display even
    shorter relaxation times, so that the approximation $V_{j,j+1}(t)=\left\langle V_{j,j+1}(t)\right\rangle_{t}$ is enough for our purposes (the hopping integrals are self-averaging).

In order to deal with the previous model, we first perform a polaron transformation of the Hamiltonian
Eq.~(\ref{eqHam}), using the generator  ${\cal U}=\exp[\sum_{\ell,\alpha}
g_{\alpha} d^{\dagger}_{\ell} d_{\ell} (B^{\dagger}_{\alpha}-B_{\alpha})]$, which is nothing else as
a shift operator of the harmonic oscillators equilibrium positions. The parameter
$g_{\alpha}=\lambda_{\alpha}/\Omega_{\alpha}$ gives an effective measure of the electron-vibron
coupling strength. Since we will work in the wide-band limit for the electrode self-energies, the
renormalization of the tunneling Hamiltonian by a vibronic operator will be neglected. As a result,
 we obtain a Hamiltonian with decoupled electronic and vibronic parts and where the onsite energies are
  shifted as $\left\langle \epsilon_{j}\right\rangle_{t}\to \left\langle \epsilon_{j}\right\rangle_{t} - \int^{\infty}_{0}d\omega J(\omega)/\omega$.
  However, as it is well-known,~\cite{mahan} the retarded Green function of the system is now an entangled
  electronic-vibronic object that can not be treated exactly; we thus decouple it in the approximate
  way:~\cite{galperin2006,gutierrez06}
\begin{eqnarray}
\label{polaron1}
 {\cal G}_{nm}(t,t')&=&-\ii \theta(t-t') \left\langle \left[ d_{n}(t){\cal X}^{\dagger}(t),d^{\dagger}_{m}(t'){\cal X}(t')\right]_{+} \right\rangle  \\ &\approx& -\ii \theta(t-t')\left\lbrace  \left\langle d_{n}(t)d^{\dagger}_{m}(t')\right\rangle  \left\langle {\cal X}^{\dagger}(t) {\cal X}(t') \right\rangle +  \left\langle d^{\dagger}_{m}(t')d_{n}(t)\right\rangle  \left\langle {\cal X}(t') {\cal X}^{\dagger}(t) \right\rangle\right\rbrace \nonumber \\
&=& \theta(t-t') \left\lbrace G^{>}_{nm}(t,t') e^{-\phi(t-t')} -  G^{<}_{nm}(t,t') e^{-\phi(t'-t)}   \right\rbrace \nonumber \\
\phi(t)&=& \sum_{\alpha} (\frac{\lambda_{\alpha}}{\Omega_{\alpha}})^2 \left[ (1+N_{\alpha})e^{-\ii\Omega_{\alpha}t} + N_{\alpha}e^{+\ii\Omega_{\alpha}t} \right] \nonumber
\end{eqnarray}
In this equation, $\theta(t-t')$ is the Heaviside function and the pure bosonic operator ${\cal X}(t)=\exp[\sum_{\alpha}
g_{\alpha}(B^{\dagger}_{\alpha}-B_{\alpha})]$. In the last row of Eq.~(\ref{polaron1})
we can pass to the continuum limit and express $\phi(t)$ in terms of the bath spectral density
$J(\omega)$:~\cite{weiss_book}
\begin{eqnarray}
\label{phi}
 \phi(t)=\frac{1}{\hbar}\int^{\infty}_{0} d\omega \,
 \frac{J(\omega)}{\omega^2} \coth{\frac{\hbar\omega}{k_{\rm B}T}}(1-\cos{\omega\,t})-\ii \frac{1}{\hbar}\int^{\infty}_{0} d\omega \frac{J(\omega)}{\omega^2} \sin{\omega\,t}.
\end{eqnarray}

\subsection{The electrical current}

We derive in this section the expression we are going to use to calculate the
electrical current through the  DNA oligomer under study. Starting point
is the well-known Meir-Wingreen expression for the current from lead $s$:~\cite{mw92}
\begin{eqnarray}
 I_{s}=\frac{2e}{\hbar}\int \frac{dE}{2\pi}\, \textrm{Tr}\left\lbrace \Sigma^{<}_{s}(E)G^{>}(E) -  \Sigma^{>}_{s}(E)G^{<}(E)\right\rbrace. \nonumber
\end{eqnarray}
Now, we can exploit the decoupling approximation used in Eq.~(\ref{polaron1}) together with
the wide-band limit in the electrode-molecule coupling to write \eg, for the left electrode:
\begin{eqnarray}
 \Sigma^{<}_{\textrm L}(E)G^{>}(E)&=&\int \frac{dE'}{2\pi}\,f_{\textrm L}(E)(1-f_{\textrm R}(E'))  t(E') \Phi(E-E'), \nonumber \\
\Sigma^{>}_{\textrm L}(E)G^{<}(E)&=&\int \frac{dE'}{2\pi}\,(1-f_{\textrm L}(E))f_{\textrm R}(E')  t(E') \Phi(E'-E), \nonumber \\
\Phi(E)&=& \int \frac{dt}{\hbar} \, e^{\frac{\ii}{\hbar} E t} e^{-\phi(t)}. \nonumber
\end{eqnarray}
Hereby we have used the explicit expressions for the greater- and lesser-Green functions:
 \begin{eqnarray}
{\bf G}^{>,<}_{0}(E)={\bf G}_{0}(E)
 ({\bf \Sigma}^{>,<}_{\textrm{L}}(E)+{\bf \Sigma}^{>,<}_{\textrm{R}}(E)){\bf G}_{0}^{\dagger}(E),
 \end{eqnarray}
 the index 0  indicating
that the vibrational degrees of freedom have already been decoupled.
The transmission-like function $t(E)$ is given by
$t(E)={\textrm {Tr}} \left\lbrace {\textbf G}_{0}(E) {\bf \Gamma}_{\textrm L} {\textbf G}^{\dagger}_{0}(E)  {\bf \Gamma}_{\textrm R} \right\rbrace$.
 The retarded matrix Green function ${\textbf G}_{0}(E)$ is calculated without electron-bath coupling but including the interaction with the electrodes:
  ${\textbf G}^{-1}_{0}(E)=E+\ii \eta - {\textbf H}_{0}+\ii {\bf \Gamma}_{\textrm L} + \ii {\bf \Gamma}_{\textrm R}$, ${\textbf H}_{0}$ being the electronic part of the Hamiltonian of Eq.~(\ref{eqHam}).
  Using these  results, the right-going current can be written as
\begin{eqnarray}
\label{curr}
 I_{\textrm L}=\frac{2e}{\hbar} \int \frac{dE}{2\pi}\int \frac{dE'}{2\pi} \, t(E')\, \left\lbrace f_{\textrm L}(E)(1-f_{\textrm R}(E'))\Phi(E-E')-   (1-f_{\textrm L}(E))f_{\textrm R}(E') \Phi(E'-E)\right\rbrace, 
\label{curr0}
\end{eqnarray}
a similar expression holding for the left-going current, when the indices L and R are interchanged. The total current can be written in a  symmetrized way: $I_{\textrm T}=(I_{\textrm L}-I_{\textrm R})/2$.
By looking
at Eq.~(\ref{phi}), two limiting cases can  immediately be obtained:
the zero charge-bath coupling ($\phi=0$) which implies
 $\Phi(E)=2\pi\delta(E)$. In this limit we recover the conventional expression for coherent transport, involving only the
 transmission function $t(E)$. In the high-temperature and/or strong coupling limit to the bath, a short-time
 expansion of $\phi(t)$ can be performed, yielding:
 \begin{eqnarray}
 \phi(t)&=&\frac{t^2}{2\hbar}\int^{\infty}_{0} d\omega \,
 J(\omega) \coth{\frac{\hbar\omega}{k_{\rm B}T}}-
 \ii \frac{t}{\hbar}\int^{\infty}_{0} d\omega \frac{J(\omega)}{\omega}, \nonumber \\
 &=&\kappa_{\textrm{therm}}t^2 -\ii  \frac{E_{\textrm{reorg}}}{\hbar}t.\nonumber
\end{eqnarray}
In the former expression $\sqrt{\kappa_{\textrm{therm}}}$ is related to an inverse decoherence time. In the high temperature limit  $\kappa_{\textrm{therm}}\sim k_{\rm B}TE_{\textrm{reorg}}$. The Fourier transform of the previous expression can be calculated straightforward and gives:
 \begin{eqnarray}
 \Phi(E)=\sqrt{\frac{\pi}{\hbar^{2}\kappa_{\textrm{therm}}}}\exp{[-\frac{ (E+E_{\textrm{reorg}})^{2}} {4\hbar^{2}\kappa_{\textrm{therm}}}]}. \nonumber
 \end{eqnarray}
Thus, the current calculated using  Eq.~(\ref{curr}) becomes a convolution of $t(E)$ with a Gaussian function. Notice the similarity of this expression with a Frank-Condon factor appearing in the Marcus electron transfer theory. 

\subsection{Spectral density from molecular dynamics}
The next issue at stake is how to relate the bath spectral density $J(\omega)$ to the time series of the electronic parameters as obtained through the molecular dynamic simulations.
To illustrate this point we will consider the simple case of a single level,
whose site energy is a Gaussian random variable,   coupled to left and right electrodes
according to the Hamiltonian (we assume for simplicity that $\left\langle \epsilon(t)\right\rangle =0$):
\begin{eqnarray}
 H=\delta\epsilon(t) d^{\dagger}d+ H_{\rm tunnel} + H_{\rm leads}
\end{eqnarray}
To deal with this problem we may use equation of motion techniques for the retarded dot Green function $G(t,t')=-(\ii/\hbar) \theta(t-t') \left\langle \left\lbrace d(t),d^{\dagger}(t') \right\rbrace \right\rangle$ in the time domain:
\begin{eqnarray}
 (\ii \hbar\partial_{t} -\delta\epsilon(t)) G(t,t')=\delta(t-t')+\sum_{{\bf k},\alpha} t^{\dagger}_{{\bf k},\alpha} G_{{\bf k}\alpha}(t,t'),
\end{eqnarray}
with
$G_{{\bf k}\alpha}(t,t')= -(\ii/\hbar) \theta(t-t')\left\langle \left\lbrace c_{{\bf k},\alpha}(t),d^{\dagger}(t')\right\rbrace  \right\rangle$. A similar equation of motion for the latter function allows to introduce the electrode self-energies:
\begin{eqnarray}
 &&(\ii\hbar \partial_{t} -\delta\epsilon(t)) G(t,t')=\delta(t-t')+\int d\tau \Sigma(t,\tau)G(\tau,t'), \nonumber \\
&&\Sigma(t,\tau)=\sum_{{\bf k},\alpha} |t_{{\bf k},\alpha}|^2 e^{-\ii\epsilon_{{\bf k},\alpha} t}
\end{eqnarray}

Note that  the Green function  thus obtained is still a random function, since no average over the
random variable $\delta\epsilon(t)$ has been yet performed.
Using the wide-band limit in the lead self-energies $\Sigma_{\alpha}(t,t')=-\ii \Gamma_{\alpha}\delta(t-t')$
we get, with $\Gamma=\Gamma_{L}+\Gamma_{R}$:
\begin{eqnarray}
 (\ii\hbar \partial_{t}  -\delta\epsilon(t)+\ii\Gamma) G(t,t')=\delta(t-t').
\end{eqnarray}
Introducing  the time-exponential ${\cal U}(t,t')=\exp(-(\ii/\hbar)\int^{t}_{t'} ds (\delta\epsilon(t)-\ii\Gamma)$ allows to write a closed solution for the dot's Green function:
\begin{eqnarray}
 G(t,t')=-\frac{\ii}{\hbar} \theta(t-t') {\cal U}(t,t')
\end{eqnarray}
The average Green function over the random variable $\delta\epsilon(t)$ yields now in energy-space:
\begin{eqnarray}
\left\langle  G(E) \right\rangle =-\frac{\ii}{\hbar} \int^{\infty}_{0} dt \, e^{\frac{\ii}{\hbar} E (t-t')}  \left\langle {\cal U}(t-t') \right\rangle = -\ii \int_{0}^{\infty} dt e^{\frac{\ii}{\hbar}(E+\ii\Gamma)(t-t')} \left\langle e^{-\frac{\ii}{\hbar}\int^{t}_{t'} ds \delta\epsilon(s)} \right\rangle.
\end{eqnarray}
Performing now a cumulant expansion~\cite{kubo62} of the averaged exponential and taking into account that cumulants higher than the second one exactly vanish due to the Gaussian nature of the fluctuations and to the fact that $\delta\epsilon(t)$ is a classical variable, we get (with $t'=0$):
 \begin{eqnarray}
\left\langle  G(E) \right\rangle =-\frac{\ii}{\hbar} \int_{0}^{\infty} dt\, e^{\frac{\ii}{\hbar}(E+\ii\Gamma)t} \, e^{-\frac{1}{\hbar^2}\int^{t}_{0} ds \int^{s}_{0} ds'\, \left\langle \delta\epsilon(s) \delta\epsilon(s') \right\rangle},
\label{stochastic}
\end{eqnarray}
which is the formally exact solution of the problem.
We may now look at the same problem from a different point of view by considering {\it explicitly} the coupling of a single site with time-independent onsite energy to a continuum of vibrational excitations. Using the polaron transformation together with
the approximations introduced at the beginning of Sec.~{\bf {II B}}, we can write the retarded Green function as:
 \begin{eqnarray}
\label{bath}
G(E)=-\frac{\ii}{\hbar} \int_{0}^{\infty} dt\, e^{\frac{\ii}{\hbar}(E+\ii\Gamma)t} \, e^{-\phi(t)},
 \end{eqnarray}
where $\phi(t)$ has been already defined in Eq.~(\ref{phi}). By comparison of Eqs.(~\ref{stochastic}) and (\ref{bath}),
 there must exist a relation between the (real) correlation function and the real part of $\phi(t)$. The latter can be re-written in the following way:
\begin{eqnarray}
  {\textrm {Re}}\, \phi(t)&=&\frac{1}{\hbar}\int^{\infty}_{0} d\omega \, \frac{J(\omega)}{\omega^2} \coth{\frac{\hbar\omega}{k_{\rm B}T}}(1-\cos{\omega\,t}) \nonumber \\&=&
\int^{t}_{0} ds\, \int^{s}_{0} ds' \left\lbrace \frac{1}{\hbar}\int^{\infty}_{0} d\omega \, J(\omega) \coth{\frac{\hbar\omega}{k_{\rm B}T}} \cos{[\omega\,(s-s')]} \right\rbrace  \nonumber,
 \end{eqnarray}
so that from this we conclude that
\begin{eqnarray}
 \left\langle \delta\epsilon(s) \delta\epsilon(s') \right\rangle=\hbar\int^{\infty}_{0} d\omega \, J(\omega) \coth{\frac{\hbar\omega}{k_{\rm B}T}} \cos{[\omega\,(s-s')]}. \nonumber
\end{eqnarray}
From here it follows by inversion:
\begin{eqnarray}
\label{spec}
J(\omega)= \frac{2}{\pi\hbar} \tanh{\frac{\hbar\omega}{k_{\textrm B} T}} \int_{0}^{\infty} dt\, \cos{\omega t} \, C(t)=\frac{2}{\pi\hbar} \tanh{\frac{\hbar\omega}{k_{\textrm B} T}} j(\omega).
\end{eqnarray}
The previous result was obtained in Ref.~\onlinecite{PhysRevE.65.031919} in the weak coupling, perturbative regime
 to the vibrational system; we have shown here that it can be also extended to the case of arbitrary coupling. Similar expressions could be derived for (spatially) non-local spectral densities.

\subsection{Results}

 \begin{figure}[t]
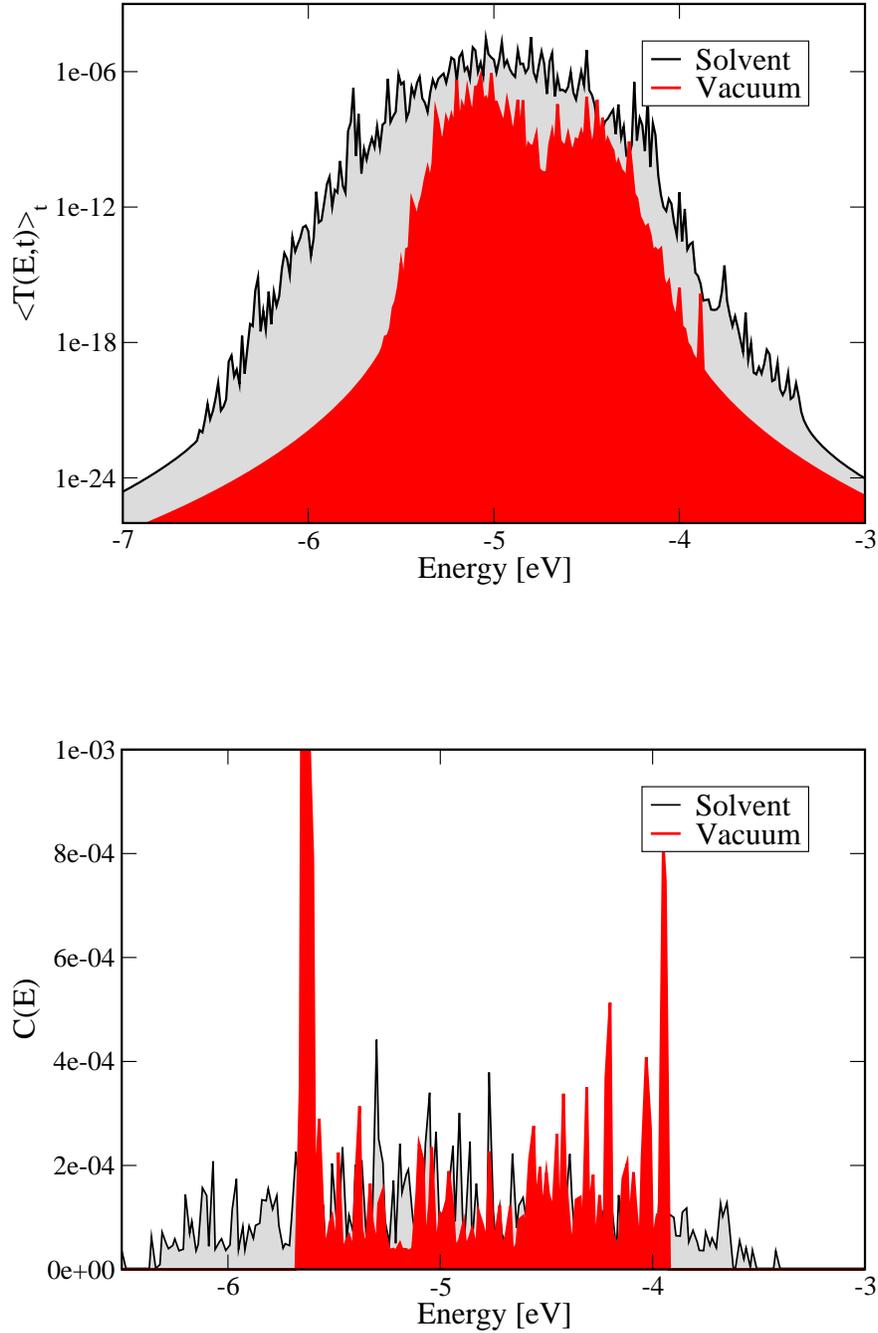

\centerline{
\epsfclipon
\includegraphics[width=0.7\linewidth]{./Figure_3a.eps}%
\epsfclipoff
}
\vfill
\vspace{2cm}
\centerline{
\epsfclipon
\includegraphics[width=0.7\linewidth]{./Figure_3b.eps}%
\epsfclipoff
}
\caption{\label{fig:corr}%
Upper panel: Time averaged transmission function for simulations performed in solvent and in vacuum. Lower panel: Coherence parameter $C(E)$ for the Dickerson dodecamer in solvent and in vacuum. Notice that the influence of the solvent fluctuations is to spread out the spectral support of $C(E)$. As expected, the coherence parameter is on average larger for the vacuum case where a large part of the fluctuations is suppressed due to the absence of an environment. Nevertheless the transmission in the vacuum case is much smaller, nicely illustrating the positive influence of the environment in gating charge migration. 
}
\end{figure}

\begin{figure}[t]
\centerline{
\epsfclipon
\includegraphics[width=0.99\linewidth]{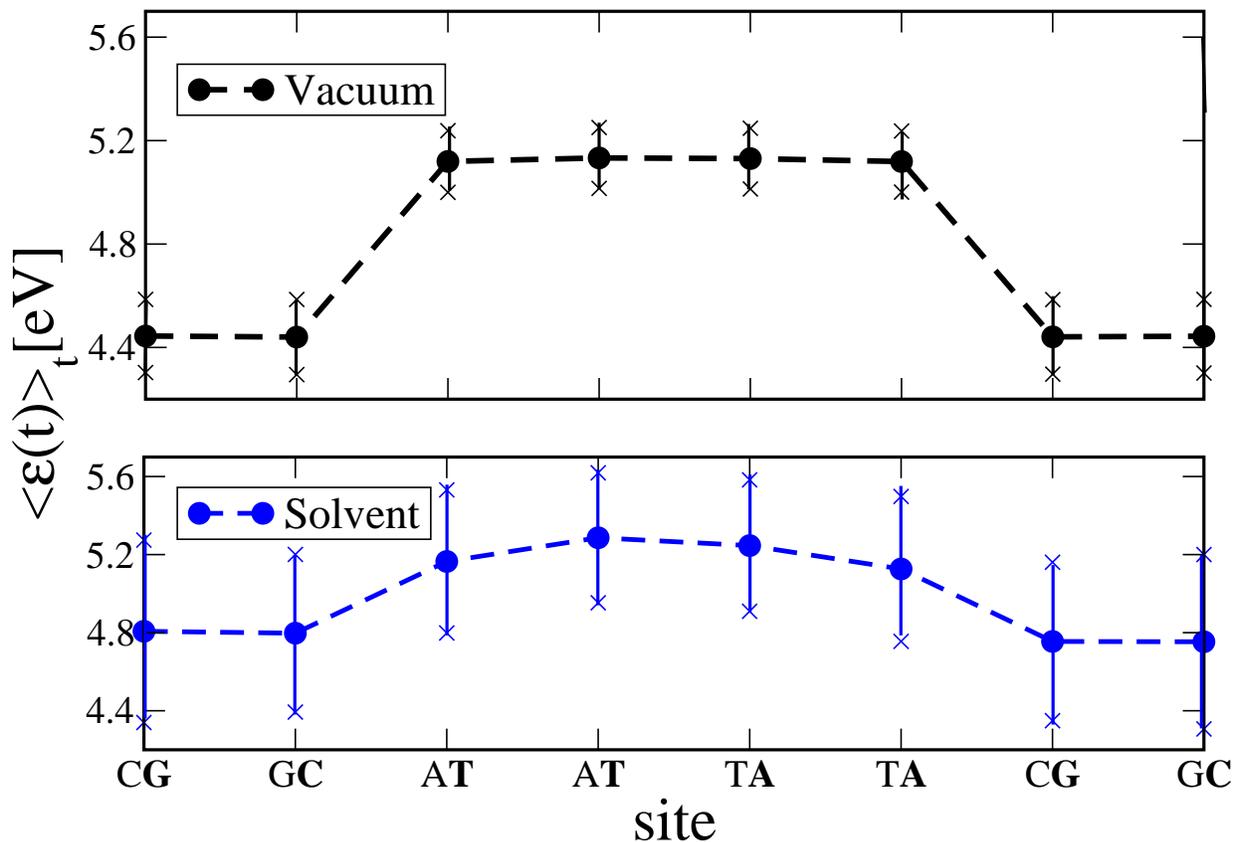}%
\epsfclipoff
}
\caption{\label{fig:onsite}%
Time averaged of the absolute value of the onsite energies along the segment covering  bases 3 and 10 of the Dickerson dodecamer, \ie, over the sequence 
 $3^{'}-$GC{\textbf{GCTTAACG}}GC$-5^{'}$) for simulations in vacuum (upper panel) and in solvent (lower panel). The outer most two bases on each end were not included in the calculations  to avoid undesired boundary effects. The averaged energy profile in presence of the solvent becomes smoother but also the fluctuations around the averages are stronger. This smoothing reduces the energy barriers between the sites and hence favours charge migration. 
}
\end{figure}

\begin{figure}[t]
\centerline{
\epsfclipon
\includegraphics[width=0.99\linewidth]{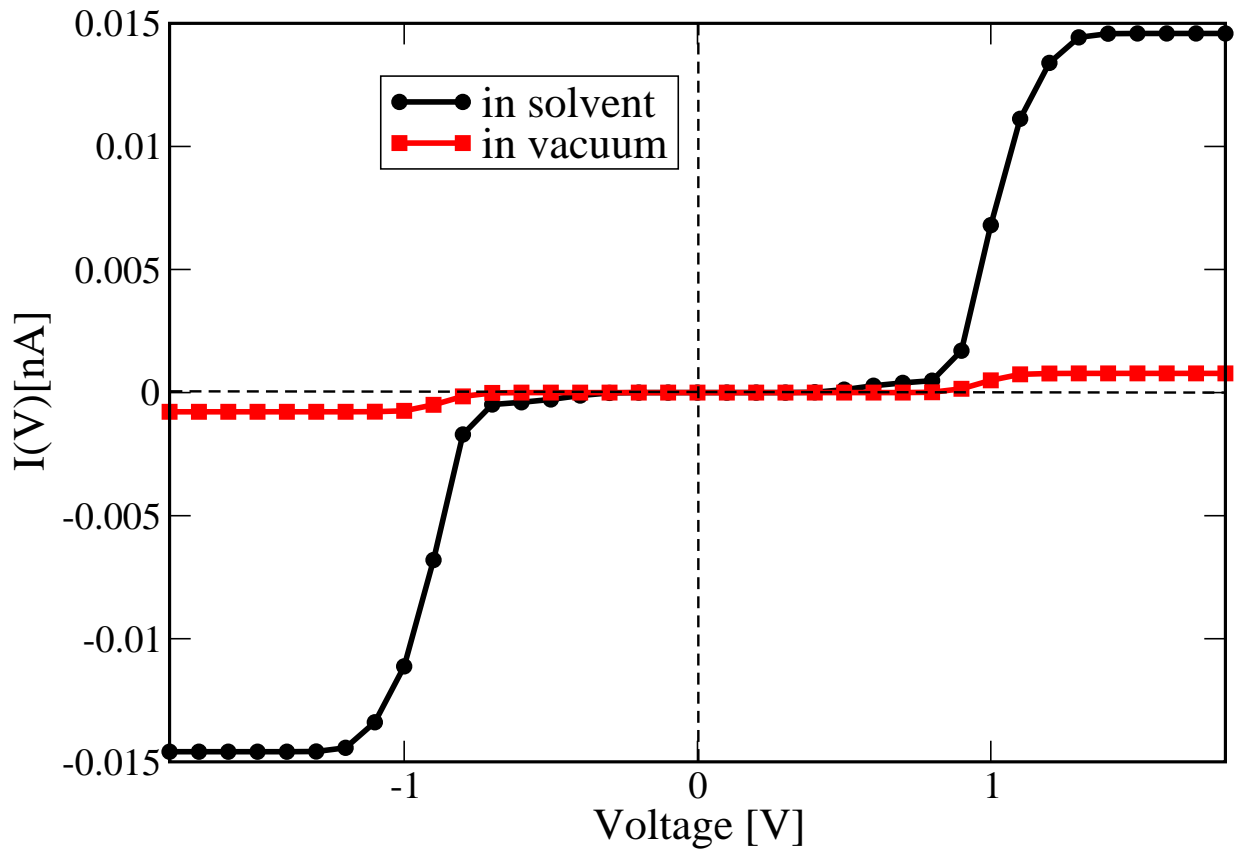}%
\epsfclipoff
}
\caption{\label{fig:curr_vac}%
Electrical current for seven base pairs of poly(dG)-poly(dC) and poly(dA)-poly(dT) 
oligomers in vacuum. The current is considerably suppressed when comparing with the case
 where solvent fluctuations are included.
}
\end{figure}

To illustrate our methodology we have focused on the Dickerson dodecamer (DD) which has a non-homogeneous base sequence. In contrast to our previous study~\cite{gutierrez2009} where homogeneous DNA sequences were addressed like poly(G)-poly(C) and  poly(A)-poly(T), the study of the DD nicely illustrates the role of the solvent fluctuations in gating the electronic structure. To study the effect of the conformational dynamics as well as the fluctuations of environment onto the charge transport properties, we have performed classical MD simulations  using the AMBER-parm99 force field92 with the parmBSC0 extension93 as implemented in the GROMACS~\cite{gromacs} software package. The static geometries were built with the 3DNA program~\cite{3dna} while the starting structures for the MD simulations were created using the make-na server.~\cite{nano} After a standard heating procedure followed by a 1 ns equilibration phase, we performed 30 ns MD simulations with a time step of 2 fs. The simulations were carried out in a rectangular box with periodic boundary conditions and filled with  5500 TIP3P water molecules and 22 sodium counterions for neutralization. Snapshots of the molecular structures were saved every 1 ps, for which the charge transfer parameters were calculated with the methodology described in the previous sections. When speaking of simulations in vacuum conditions, we mean that the last term in Eq.~\ref{QMMM} is left out. Upon mapping the electronic structure onto a linear chain as discussed in the former sections,  we have first computed the time and energy dependent quantum-mechanical transmission function $T(E,t)$  by evaluating for every set of charge transfer parameters, \ie, at each simulation time step $t$,  $\left\langle T(E,t)\right\rangle_{t}=(1/T_{\textrm{MD}})\sum_{j}T(E,t_{j})$. This quantity is expected to provide some qualitative insight into different factors affecting transport (solvent vs. vacuum) but its use is restricted by the fact that for longer chains inelastic, fluctuation mediated channels will play an increasing role and then a pure elastic transmission can not catch all the transport physics. In this latter case, the model Eq.~(\ref{polaron}) seems to us more appropriate since it includes the dressing of the electronic degrees of freedom by the structural fluctuations. In Fig.~\ref{fig:corr} we show the averaged transmission function (upper panel) for the cases where solvent effects are included or neglected, respectively. We clearly see that solvent-mediated fluctuations considerably increase and broaden the transmission spectrum. Its fragmented structure is simply due to the fluctuations in the onsite energies, which make the system effectively highly disordered. Another way of looking at the effect of fluctuations is via the introduction of a coherence parameter~\cite{gutierrez2009} which is defined as $C(E)=\left\langle T(E,t)\right\rangle_{t}^{2}/\left\langle T(E,t)^2\right\rangle_{t}$. If the fluctuations are weak $C(E$ goes to one, while a strong fluctuating system will lead to a considerably reduction of $C(E)$. Obviously, this parameter has only a clear meaning over the spectral support of the transmission function. $C(E)$ is shown in the lower panel of  Fig.~\ref{fig:corr}; the coherence parameter in vacuum conditions can become larger than in solvent for some small energy regions due to the reduction of the fluctuations. However, this does not turn to be enough to increase the transport efficiency for the special base sequence of the Dickerson DNA, since this system has in the static limit a distribution of energetic barriers due to the base pair sequence. To further illustrate the influence of the solvent, we have plotted in Fig.~\ref{fig:onsite}  the time averaged onsite energies along the  model tight-binding chain. Remarkably, the presence of the solvent ''smoothes'' the averaged energy profile (though the amplitude of the  fluctuations clearly becomes stronger). To compute the current through the system, we use the Hamiltonian Eq.~(\ref{polaron}) and the current expression, Eq.~(\ref{curr0}), derived with this model. The Fermi energy in these calculations was fixed at the upper edge of the transmission spectrum, see Fig.~\ref{fig:corr}. The qualitative results are however not considerably changed by slightly changing its position. In Fig.~\ref{fig:curr_vac} the current is shown for the two cases of interest.  Due to the presence of tunnel barriers in the wire which are not fully compensated on average by the gating effect of the environment, the absolute current values  are  rather small when compared with those of homogeneous sequences~\cite{gutierrez2009}. However,  the current including the solvent is roughly fifteen times larger than for that obtained from the simulations in vacuum. Though our model Hamiltonian in Eq.~(\ref{polaron}) does not fully contain all the dynamical correlations encoded in the time dependent electronic parameters, we nevertheless expect that their inclusion would lead to an even further increase of the difference between solvent and vacuum results, thus suppporting our main conclusions.

\section{Conclusions}
We have presented a hybrid methodology which allows for a very efficient coarse-graining of the electronic structure of complex biomolecular systems as well as to include the influence of structural fluctuations into the electronic parameters. The time series obtained in this way allow to map the problem onto an effective low dimensional model Hamiltonian describing the interaction of charges with a bosonic bath, which comprises the dynamical fluctuations. The possibility to parametrize the bath spectral density $J(\omega)$ using the information obtained from the time series makes our approach very efficient, since we do not need to use the typical phenomenological {\textit{Ans\"atze}} (ohmic, Lorentzian, etc) to describe the bath dynamics. The example presented here, the Dickerson dodecamer, shows in a very clear way that solvent-mediated gating may be a very efficient mechanism in supporting charge transport if the static DNA reference structure already posseses a disordered energy profile (due to the base sequence). Our method is obviously not limited to the treatment of DNA but it can equally well be applied to deal with charge migration in other complex systems like molecular organic crystals or polymers, where charge dynamics and coupling to fluctuating environments plays an important role.~\cite{citeulike:5831061,GudowskaNowak1996115,goychuk:4937,PhysRevB.3.262,SpirosS.Skourtis03082005,ADR2008,TRZ2004,bicout1995}

\section{Acknowledgments}
The authors acknowledge Florian Pump for fruitful discussions. This work has been  supported by the Deutsche Forschungsgemeinschaft (DFG) within the Priority Program 1243 ``Quantum transport at the molecular scale'' under contract CU 44/5-2, by the  Volkswagen Foundation grant Nr.~I/78-340,  by the European Union under contract IST-029192-2. We further acknowledge the Center for Information Services and High Performance Computing (ZIH) at the Dresden University of Technology for computational resources.

%

\end{document}